\documentclass[twocolumn,prl,superscriptaddress,a4paper]{revtex4}
\usepackage{amsmath,amssymb,graphicx}
\usepackage[usenames]{color}
\begin{document}

\setlength{\tabcolsep}{1.5mm}
\setcounter{totalnumber}{4}
\setcounter{topnumber}{4}

\setlength{\voffset}{-0cm}
\setlength{\hoffset}{-0.cm}
\addtolength{\textheight}{1.1cm}

%%%%%%%%%%%%%%%%%%%%
%% PRIVATE MACROS %%
%%%%%%%%%%%%%%%%%%%%

\newcommand{\figwidth}{0.90\columnwidth}
\newcommand{\eq}[1]{Eq.(\ref{#1})}
\newcommand{\fig}[1]{Fig.~\ref{#1}}
\newcommand{\sect}[1]{Sec.~\ref{#1}}
\newcommand{\avg}[1]{{\langle #1 \rangle}}
\newcommand{\olcite}[1]{Ref.~\onlinecite{#1}}

%%%%%%%%%%%%%%%%%%%%%%%%%%
%% DOCUMENT STARTS HERE %%
%%%%%%%%%%%%%%%%%%%%%%%%%%

\title{Phase diagram of $\alpha$-helical and $\beta$-sheet forming peptides}

\author{Stefan Auer}
\affiliation{Centre for Molecular Nanoscience, University of Leeds,
Leeds LS2 9JT, United Kingdom}

\author{Dimo Kashchiev}
\affiliation{Institute of Physical Chemistry, Bulgarian Academy of Sciences, Ul. Acad. G. Bonchev 11, Sofia 1113, Bulgaria}


\date{\today}

\begin{abstract}
The intrinsic property of proteins to form structural motifs such as
$\alpha$-helices and $\beta$-sheets leads to a complex phase behaviour
in which proteins can assemble into various types of aggregates
including crystals, liquid-like phases of unfolded or natively folded
proteins, and amyloid fibrils. Here we use a coarse-grained protein
model that enables us to perform Monte Carlo simulations for
determining the phase diagram of natively folded $\alpha$-helical and
unfolded $\beta$-sheet forming peptides. The simulations reveal the
existence of various metastable peptide phases. The liquid-like phases
are metastable with respect to the fibrillar phases, and there is a
hierarchy of metastability.
\end{abstract}

\maketitle

In a diagram of protein solution (e.g.,
refs\cite{asherie04,vekilov04,annunziata05,dumetz08}), the solubility
line specifies the conditions under which a protein crystal neither
grows nor dissolves and corresponds to the gas-crystal coexistence
line in a diagram of atomic system. Similarly, the gas-liquid
coexistence line for an atomic system corresponds to the liquid-liquid
separation line for a protein solution (the latter is the line of
coexistence of two separate liquid phases, a protein-rich and a
solvent-rich ones). The phase behavior described in
refs\cite{asherie04,vekilov04,dumetz08} involves native proteins, with
little or no difference in the protein conformation in the different
phases. However, depending on solvent conditions and physical
variables such as temperature, the proteins may also adopt non-native
conformations, resulting in the formation of various
aggregates. Indeed, a wide range of different proteins unrelated in
their amino acid sequence have been shown to form amyloid fibrils
which share a common characteristic cross-$\beta$ structure where
peptides form $\beta$-sheets oriented parallel to the fibril
axis\cite{chiti06,sawaya07}. Despite the importance of the effect of
protein conformational changes on the protein phase diagram, we are
aware of only one experimental diagram\cite{annunziata05} involving
this effect. It is for Human $\beta$B1-crystallin and describes the
coexistence lines between a solution and fibrillar aggregates and
between these aggregates and a liquid-like phase. On the other hand,
attempts have also been made\cite{dima02,nguyen04a} to numerically
construct phase diagrams of peptides by determining the ground state
structure of the condensed peptide phases. The resulting diagrams
however do not account for the existence of metastable phases that are
crucial to the understanding of generic aspects of the protein phase
behavior.

The present study makes use of a novel theoretical
framework\cite{hoang04} in which proteins are described as flexible
tubes. In the model used here and described in detail
elsewhere\cite{auer07} the protein backbone is represented by a
C$_{\alpha}$ chain with finite thickness. The directional hydrogen
bonding is sequence-independent and can be accounted for by an
analysis of the geometrical properties of hydrogen-bond forming
C$_{\alpha}$ atoms in protein structures listed in the Protein Data
Bank. The hydrogen-bond energy is denoted by ${\it\epsilon}$. The
hydrophobic effect between C$_{\alpha}$ atoms is captured by a
pairwise attractive square-well potential with energy
${\it\epsilon_{h}}$. As in previous studies\cite{auer07,auer08}, we
use ${\it\epsilon/\epsilon_h}=20$ in order to quantify the relative
strength of the hydrogen and the hydrophobicity-mediated
bondings. Steric constraints are implemented by a local bending
stiffness with energy ${\it\epsilon_s}$ per C$_{\alpha}$ atom. In all
our simulations the stiffness energy ${\it\epsilon_s}$ is set equal to
$0.3{\it\epsilon}$. Peptides that are weakly hydrophobic, and hence
have such a large ${\it\epsilon/\epsilon_h}$ value, have been
demonstrated\cite{saiani09} to assemble into the various types of
aggregates considered here. We investigated the phase behavior of a
simple prototype of biomolecular system consisting of 12-residue
homopeptides in an implicit aqueous solution. The simulations showed
that most of the peptides in the solution fold at least partially into
a native $\alpha$-helical structure at temperatures below the folding
temperature ${\it T_f}=0.20{\it\epsilon/k}$, and unfold at least
partially into an extended random-coil structure above it. Our
simulation model is thus pertinent to peptides with hydrogen-bond
energy ${\it\epsilon}=(1.9$ to $2.5)\times 10^{-20}$ J, because these
${\it\epsilon}$ values correspond to ${\it T_f}=276$ to $363$ K,
i.e. to peptide folding temperatures of biophysical relevance.
\begin{figure}
\begin{center}
\includegraphics[width=8.6cm]{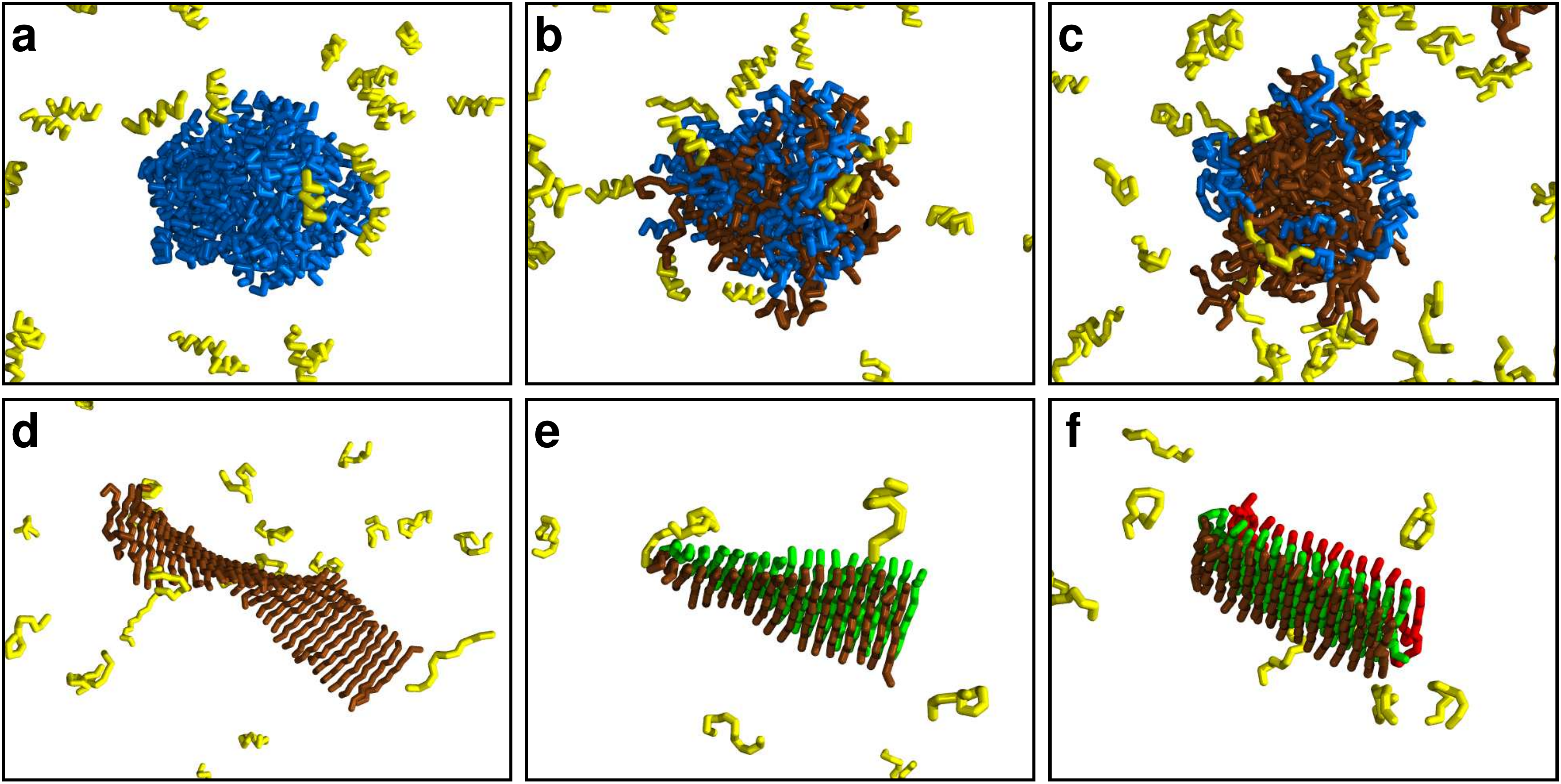}
\caption{Gallery of peptide aggregates. 
(a) $\alpha$-oligomer at dimensionless temperature ${\it
\theta\equiv T/T_f}=0.9$, (b) and (c) $\gamma$-oligomer at ${\it
\theta}=1.0$ and $1.1$, respectively, (d) $1\beta$-sheet at
${\it \theta}=1.1$, (e) $2\beta$-sheet at ${\it \theta}=1.2$, and (f)
$3\beta$-sheet at ${\it
\theta}=1.2$. Peptides in the
solution are shown in yellow, peptides within an aggregate that do not
form interpeptide hydrogen bonds are shown in blue, and those that do
so are shown in brown, green or red.\label{fig1}}
\end{center}
\end{figure}

\begin{figure}
\begin{center}
\includegraphics[width=7.5cm]{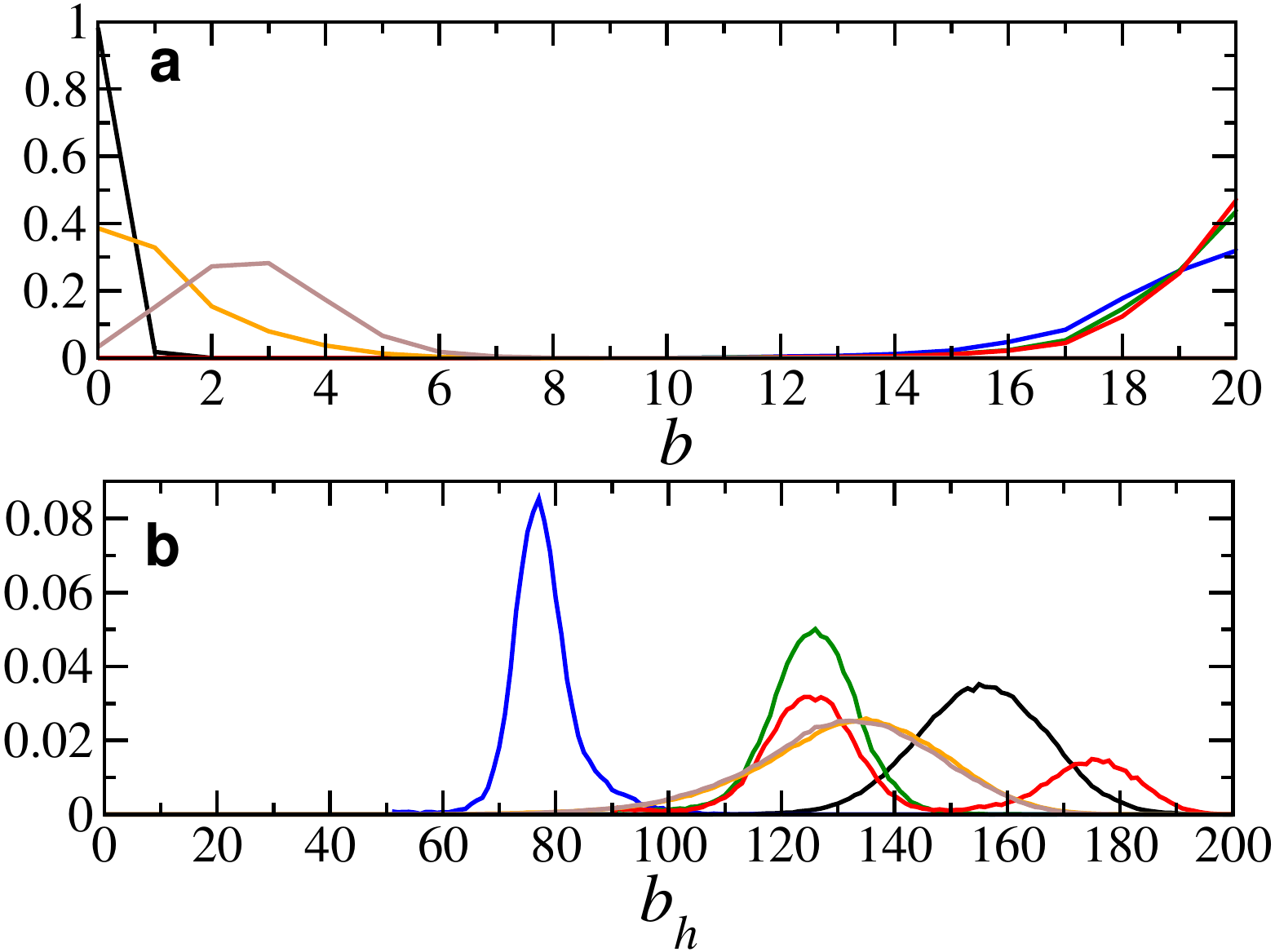}
\caption{Bonding in peptide aggregates. (a)
Probability distribution function of the number ${\it b}$ of hydrogen
bonds per peptide in: $\alpha$-oligomer (black line) at ${\it
\theta}=0.7$, $\gamma$-oligomer (orange line) at ${\it
\theta}=1.0$, $\gamma$-oligomer (brown line) at ${\it
\theta}=1.1$, $1\beta$-sheet (blue line) at ${\it \theta}=1.1$, $2\beta$-sheet 
(green line) at ${\it \theta}=1.2$, and $3\beta$-sheet (red line) at
${\it\theta}=1.2$. (b) Probability distribution function of the
number ${\it b_h}$ of hydrophobicity-mediated bonds per peptide in:
$\alpha$-oligomer (black line), $\gamma$-oligomers (orange and brown
lines), $1\beta$-sheet (blue line), $2\beta$-sheet (green line), and
$3\beta$-sheet (red line) at the respective ${\it \theta}$ values
noted above. 
\label{fig2}}
\end{center}
\end{figure}

Our simulations showed that the peptides can self-assemble into
different types of aggregates: $\alpha$-oligomers - disordered
aggregates constituted solely of fully folded $\alpha$-helical
peptides (Fig.~\ref{fig1}a), $\gamma$-oligomers - also disordered
aggregates but consisting of fully folded, partially folded and
unfolded peptides (Figs.~\ref{fig1}b and~\ref{fig1}c), and ${\it
i}\beta$-sheets - ordered aggregates with ${\it i}=1,2,3,...$ layers
of single $\beta$-sheet tapes (Figs.~\ref{fig1}d-\ref{fig1}f,
respectively). Importantly, as seen in Figs.~\ref{fig1}b
and~\ref{fig1}c, the fraction of folded/unfolded peptides in the
$\gamma$-oligomers depends on temperature. We shall consider the above
aggregates as representing distinct phases, because the relative
contribution of the hydrophobicity-mediated and the hydrogen bondings
that stabilize them is different (Fig.~\ref{fig2}). While an
$\alpha$-oligomer is stabilized by the hydrophobicity-mediated bonding
only, the $\gamma$-oligomer is additionally stabilized by the hydrogen
bonding (by definition, a $\gamma$-oligomer is completely free of
${\it i}\beta$-sheets). In contrast, the ${\it i}\beta$-sheets are
predominantly stabilized by the hydrogen bonding. On average
(Fig.~\ref{fig2}), for the number ${\it b}$ of interpeptide hydrogen
bonds per peptide we have ${\it b}=0$ for the $\alpha$-oligomers,
${\it b}=1$ or 3 for the $\gamma$-oligomers at dimensionless
temperature ${\it\theta\equiv T/T_f}=1.0$ or 1.1, respectively, and
${\it b}=18$ for all ${\it i}\beta$-sheets.  Also on average
(Fig.~\ref{fig2}), the number ${\it b_h}$ of hydrophobicity-mediated
bonds per peptide is $155$, 78, 124 and 134 for the $\alpha$-oligomer,
$1\beta$-sheet, $2\beta$-sheet and $3\beta$-sheet, respectively.  For
the $\gamma$-oligomer, ${\it b_h}=132$ at ${\it\theta}=1.0$ and 1.1.

\begin{figure}
\begin{center}
\includegraphics[width=8.cm]{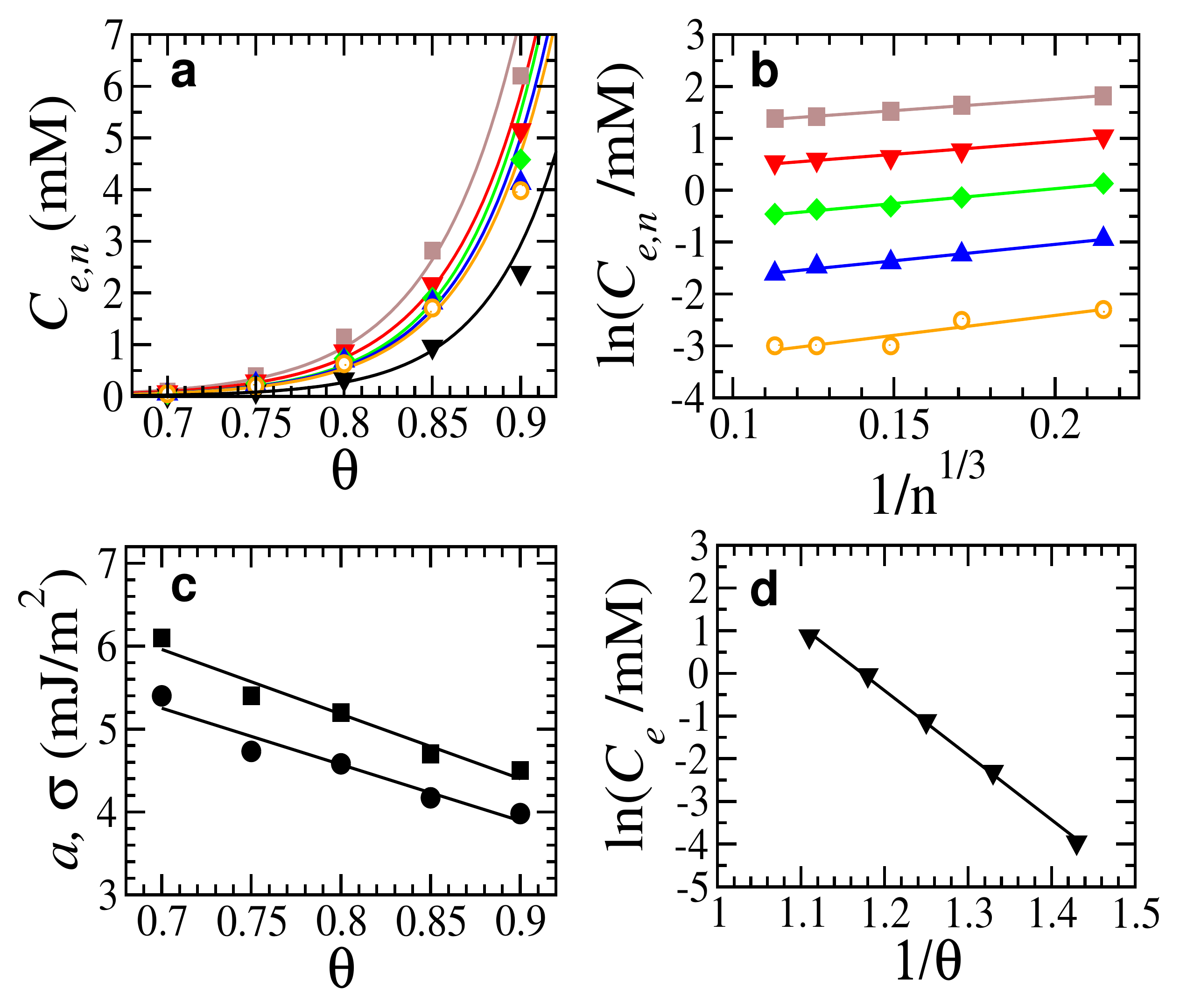}
\caption{$\alpha$-oligomer results. (a) Temperature 
dependence of solubility: squares, open down triangles, diamonds, up
triangles and circles - simulation data for cluster size ${\it
n}=100$, 200, 300, 500 and 700, respectively; black trianlges -
derived data for ${\it n}=\infty$ from best-fit analysis; lines - best
fit of equation~\ref{eq2}. (b) Size dependence of solubility:
squares, open down triangles, diamonds, up triangles and circles -
simulation data at ${\it\theta}=0.9$, 0.85, 0.8, 0.75, 0.7,
respectively; lines - best fit of equation~\ref{eq1}. (c)
Dependence of parameter ${\it a}$ (circles) and specific surface
energy ${\it\sigma}$ (squares) on ${\it\theta}$; the straight lines
are drawn to guide the eye. (d) Solubility of bulk
$\alpha$-oligomeric phase: triangles - data from (a) for ${\it
n}=\infty$; line - best fit of equation~\ref{eq2}.
\label{fig3}}
\end{center}
\end{figure}

\begin{figure}
\begin{center}
\includegraphics[width=7.5cm]{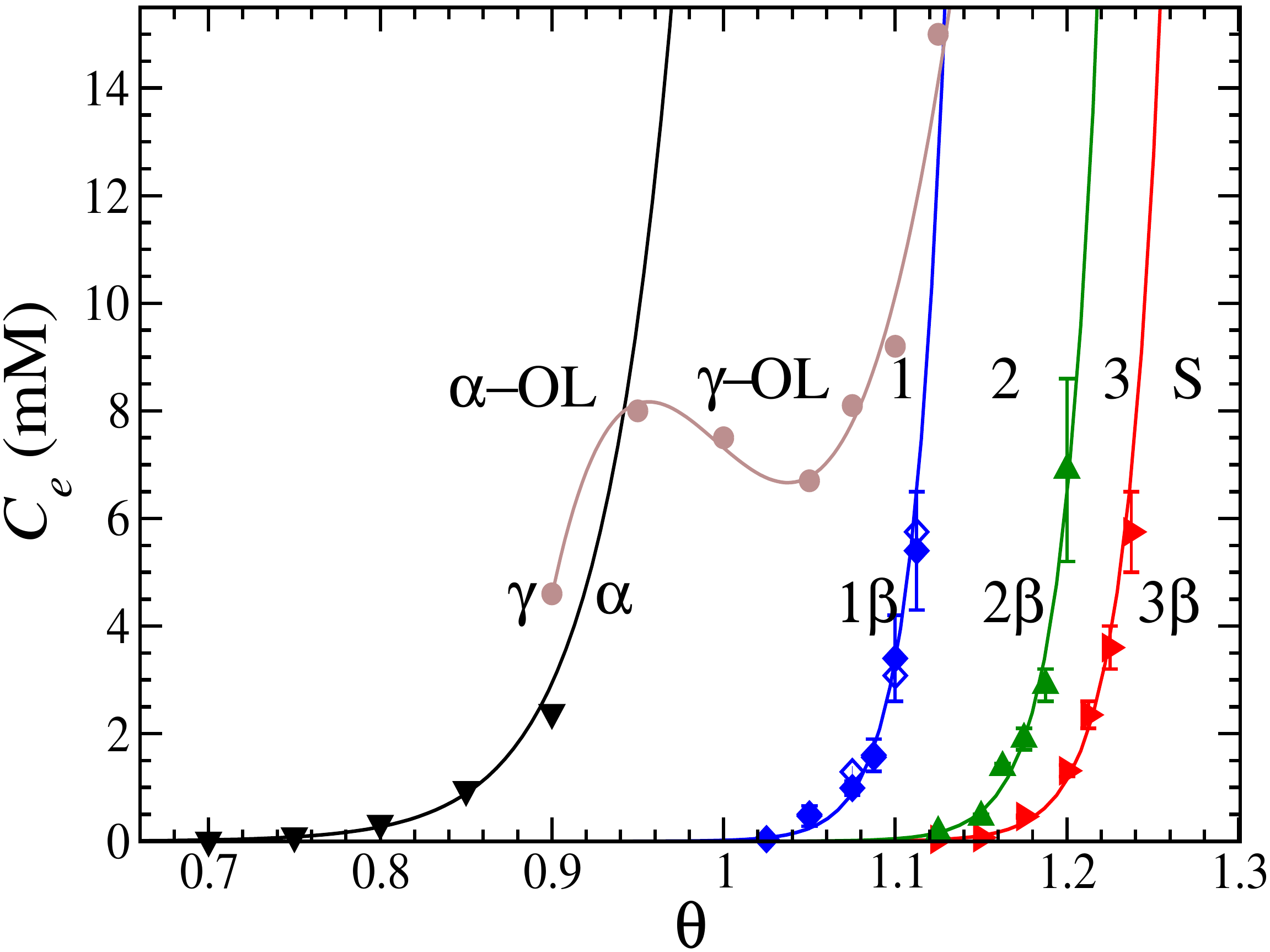}
\caption{Peptide phase diagram. Solubility-vs-temperature data for 
$\alpha$-oligomer (down triangles) of infinitely large size,
$\gamma$-oligomer (circles) of size ${\it n}=100$, $1\beta$-sheet
(solid diamonds), $2\beta$-sheet (up triangles) and $3\beta$-sheet
(right triangles), all of size ${\it n}=40$. Lines $\alpha$, $1\beta$,
$2\beta$ and $3\beta$ are obtained by best fit of equation~\ref{eq2}
to the data; line $\gamma$ is just a guide to the eye; labels
$\alpha$-OL, $\gamma$-OL, 1, 2, 3, and S refer to different stability
regions (see main text); the open diamonds refer to ${\it
n}=20$. 
\label{fig4}}
\end{center}
\end{figure}

In order to determine the solubility lines for the $\alpha$-oligomer,
$\gamma$-oligomer and ${\it i}\beta$-sheets with ${\it i}=1,2,3$ we
followed the approach used by Bai and Li\cite{bai06}. We performed
Monte Carlo simulations in the canonical ensemble using crankshaft,
pivot, reptation, rotation and translation moves. The simulations were
carried out by using a cubic box and periodic boundary conditions. For
the $\alpha$-oligomer solubility determination we prepared a cluster
of a fixed number ${\it n}$ of fully folded peptides and placed this
${\it n}$-sized cluster into a solution containing also a fixed number
(100 or 200) of such peptides. 
Isothermally, we performed several simulation runs at different
peptide concentrations and monitored whether the cluster grew or
shrank. As in this case the cluster was an $\alpha$-oligomer, we
constrained the peptides in the system to remain fully folded by
performing only rotation and translation moves during the
simulation. This enabled us to calculate the $\alpha$-oligomer
solubility at temperatures at which the peptides would normally start
to unfold. From the first simulation runs we identified two near
peptide concentrations between which the $\alpha$-oligomer would
coexist with the solution. This provided a concentration range for
additional runs which yielded more accurately the equilibrium
concentration (or solubility) ${\it C_{e,n}}$ at which an ${\it
n}$-sized $\alpha$-oligomer neither grows nor shrinks. As cluster
solubility is known to depend on cluster size, we studied the
dependence of ${\it C_{e,n}}$ on ${\it n}$ at a given ${\it\theta}$ by
repeating our simulations for five $\alpha$-oligomer sizes: ${\it n}=
100$, 200, 300, 500 and 700. The results obtained are shown in
Fig.~\ref{fig3}a.
The size dependence of ${\it C_{e,n}}$ can be
described by the Ostwald formula (e.g., ref.\cite{kashchiev00})
\begin{equation}
{\it 
\ln C_{e,n} = \ln C_e} + {\it a}/{\it \theta n}^{1/3}
\label{eq1}
\end{equation}
in which ${\it C_e}$ is the equilibrium concentration (or solubility)
of the infinitely large $\alpha$-oligomer, and ${\it a=2c\sigma
v}^{2/3}/3{\it kT_f}$. Here ${\it\sigma}$ is the specific surface
energy of the $\alpha$-oligomer/solution interface, ${\it v}$ is the
volume occupied by a peptide in $\alpha$-oligomer, and for a spherical
$\alpha$-oligomer the shape factor ${\it c}=(36{\it\pi})^{1/3}$. In
$\ln{\it C_{e,n}}$-vs-$1/{\it n}^{1/3}$ coordinates, a linear fit of
equation~\ref{eq1} to our data (Fig.~\ref{fig3}b) yielded values for
${\it \ln C_e}$ (from the intercept) at each temperature
${\it\theta}$, and hence the solubility ${\it C_e(\theta)}$ of the
infinitely large $\alpha$-oligomer (the down triangles in
Fig.~\ref{fig3}a and in Fig.~\ref{fig4}). Also, the slope of the fit
provided the ${\it a}$ value at each ${\it\theta}$. The resulting
${\it a(\theta)}$ dependence (the circles in Fig.~\ref{fig3}c) implies
that the $\alpha$-oligomer surface tension ${\it\sigma}$ decreases
with temperature as illustrated by the squares in
Fig.~\ref{fig3}c. These squares represent the ${\it\sigma}$ values
calculated from the above expression for ${\it a}$ with the aid of
${\it c}=(36{\it\pi})^{1/3}$ (spherical $\alpha$-oligomers), ${\it
T_f}=300$ K (peptides with ${\it\epsilon}=2.07\times 10^{-20}$ J) and
${\it v}=1.2$ nm$^3$. The latter is the mid-range value of the peptide
volume for ${\it\theta}$ between 0.7 and 0.9, and was obtained by
dividing the volume of all peptides in the bulk of an
$\alpha$-oligomer by the number of these peptides. As seen in
Fig.~\ref{fig3}c, ${\it\sigma}= 4.5$ to 6.1 mJ/m$^2$ and is in the
range of 0.1 to 30 mJ/m$^2$, values reported both theoretically and
experimentally for the specific surface energy of protein crystals in
aqueous solutions\cite{haas95,chernov03,vekilov04}.

Fitting the integrated van't Hoff equation
\begin{equation}
{\it 
C_e=C_r\exp(-\lambda/\theta) }
\label{eq2}
\end{equation}
to the ${\it C_{e}(\theta)}$ data in Fig.~\ref{fig3}d enables
determining the dimensionless latent heat ${\it \lambda=L/kT_f}$ of
peptide aggregation into an infinitely large $\alpha$-oligomer. Here
${\it C_r}$ is a practically temperature-independent reference peptide
concentration, and ${\it L}$ is the latent heat of peptide aggregation
into such an oligomer. The best-fit result is ${\it
C_r}=5.2\times 10^{7}$ mM and ${\it\lambda}=15.1\pm 0.3$. With ${\it
T_f}=0.2{\it\epsilon/k}$ it thus follows that ${\it L=\lambda k T_f}$
has the value of $3{\it\epsilon}$. Hence, recalling that in our
simulations ${\it\epsilon}=20{\it\epsilon_h}$ and that on average
${\it b_h}=155$, we find that the latent heat ${\it
L}=60{\it\epsilon_h}$ of peptide aggregation into $\alpha$-oligomer is
somewhat lower (by the factor 0.78)
than half of the average binding energy ${\it
b_h}\epsilon_h=155{\it\epsilon_h}$ that a fully folded peptide has in
the bulk $\alpha$-oligomeric phase. For comparison, in the Haas-Drenth
lattice model\cite{haas95} of protein crystals the latent heat of
crystallisation is just half the average protein binding energy. Also,
with ${\it a}=4.0$ to $5.4$ (see Fig.~\ref{fig3}c),
${\it\lambda}=15.1$ and ${\it c}=(36{\it\pi})^{1/3}$, for the ratio
${\it\sigma v}^{2/3}/{\it L}=3{\it a}/2{\it c\lambda}$ we obtain
values from $0.08$ to $0.11$ in the $\theta$ range studied. Thus,
${\it\sigma v}^{2/3}/{\it L}$ for the $\alpha$-oligomers is
considerably smaller than for atomic or simple molecular substances
which are known to follow the Stefan-Skapski-Turnbull relation
${\it\sigma v}^{2/3}/{\it L}=0.2-0.6$ (e.g., Ref.~\cite{kashchiev00}).

Next in our study was the determination of the $\gamma$-oligomer
solubility. Following the procedure outlined for the
$\alpha$-oligomers, we prepared a cluster of size ${\it n}=100$. As
the $\gamma$-oligomer consists of fully folded, partially folded and
unfolded peptides, their proportion in it was taken to be the same as
that of the peptides in the solution at the chosen temperature. The
simulations, in which we performed pivot, crankshaft, reptation,
rotation and translation moves, revealed that the $\gamma$-oligomer
was a transient formation always transforming into an ${\it
i}\beta$-sheet. For that reason it was necessary to use a biasing
potential that arrested the $\gamma$-oligomer transformation by not
allowing a peptide to form more than one interpeptide hydrogen bond
with any other peptide in the cluster. The simulation ${\it
C_{e,n}(\theta)}$ data for $\gamma$-oligomers of size ${\it n}=100$
(the circles in Fig.~\ref{fig4}) show that the $\gamma$-oligomer
solubility cannot be described by equation~\ref{eq2}. The
non-monotonic behavior of the $\gamma$-oligomer solubility is a result
of the temperature dependence of the proportion of folded and unfolded
peptides in the oligomer. This effect is strongest for $\theta$
between 0.9 and 1.1, the temperature range in which the peptides
commence unfolding and forming additional interpeptide hydrogen bonds
that decrease the $\gamma$-oligomer solubility. Partial unfolding of
proteins is regarded as being a crucial step in protein
aggregation\cite{uversky04,platt05} and our observation of the
decreased $\gamma$-oligomer solubility with the peptide unfolding
rationalizes this view. Regrettably, as the simulations with
$\gamma$-oligomers were computationally much more demanding than those
with the $\alpha$-oligomers, we could not investigate the effect of
the cluster size on the $\gamma$-oligomer solubility.

Finally, we determined the solubility lines for the $1\beta$-,
$2\beta$- and $3\beta$-sheets. In these simulations all ${\it
i}\beta$-sheets consisted of ${\it n}=40$ unfolded peptides. Since
${\it i}\beta$-sheet formation at a fixed ${\it i}$ value is a
one-dimensional clustering process, the ${\it i\beta}$-sheet
solubility is expected to be ${\it n}$-independent. We could not
confirm this expectation for all ${\it i}=1,2$ and $3$, but
simulations with $1\beta$-sheets of size ${\it n}=20$ revealed that
these sheets indeed had the solubility of those with ${\it n}=40$ (see
the diamonds in Fig.~\ref{fig4}). The simulations showed also that a
new peptide monolayer could occasionally form on the ${\it
i}\beta$-sheet studied, transforming it into an ${\it (i }
+1)\beta$-sheet. In order to calculate the solubility of these
metastable ${\it i}\beta$-sheets we had to include a biasing potential
preventing this transformation from occurring. The so-obtained
simulation ${\it C_{e,n}(\theta)}$ data are shown in Fig.~\ref{fig4}
by symbols. A fit of equation~\ref{eq2} to these data provides values
for the reference peptide concentration and latent heat of peptide
aggregation into an ${\it i}\beta$-sheet: ${\it C_r}=2.1\times 10^{24}$
mM and ${\it\lambda}=60\pm 6$ for the $1\beta$-sheet, ${\it C_r}=4.3\times
10^{24}$ mM and $\lambda= 66\pm 4$ for the $2\beta$-sheet, and
$C_r=1.2\times 10^{24}$ mM and ${\it\lambda}=66\pm 3$ for the
$3\beta$-sheet. With the ${\it T_f}$ value given above, it thus
follows that ${\it L=\lambda k T_f}$ has the values of $12\epsilon$,
$13.2\epsilon$ and $13.2\epsilon$ for the $1\beta$-, $2\beta$- and
$3\beta$-sheet, respectively. Again, these ${\it L}$ values can be
compared with the average binding energy ${\it b\epsilon +
b_h\epsilon_h} = 22{\it \epsilon}$, $24{\it
\epsilon}$ and $24{\it \epsilon}$ that an unfolded peptide has in the
bulk $1\beta$-, $2\beta$- and $3\beta$-sheet, respectively. These
numbers follow from the ${\it b}$, ${\it b_h}$ and
${\it\epsilon/\epsilon_h}$ values already given above. We see that,
similar to the $\alpha$-oligomer, and in close correspondence with
lattice models, for the ${\it i}\beta$-sheets ${\it L}$ is about half
the peptide average binding energy. We note that the ${\it L}$ value
for the $3\beta$-sheet may be regarded as representative for that of
the infinitely thick $\beta$-sheet, because in our simulation model
the hydrophobicity-mediated interaction is limited within two
successive peptide monolayers only. This limitation is reflected by
the virtually equal average binding energies of a peptide in the
$2\beta$- and $3\beta$-sheets. Also, the ${\it L}$ values for the
${\it i}\beta$-sheets are considerably greater than those for the
$\alpha$- and $\gamma$-oligomers because of the presence of more
hydrogen bonds per peptide in the ${\it i}\beta$-sheets.

In conclusion, the phase diagram in Fig. 4 shows that there are two
regions of thermodynamic stability, one of the peptide solution
(region S), and one of the infinitely thick $\beta$-sheet. The
dividing line between these two stable phases is approximately the
solubility line of the $3\beta$-sheet (line $3\beta$), because in our
model the peptide interactions are restricted within two successive
$\beta$-sheets. In addition to this, the phase diagram reveals the
hierarchic existence of various metastable peptide phases. In region
$3$ the $3\beta$-sheet is stable with respect to the solution, but
metastable with respect to thicker ${\it i}\beta$-sheets ($i>3$). In
region $2$ the $2\beta$-sheet is stable with respect to the solution,
but metastable with respect to all thicker $\it{i}\beta$-sheets
($i>2$). In region $1$ the $1\beta$-sheet is stable with respect to
the solution, but metastable with respect to all other ${\it
i}\beta$-sheets ($i>1$). In region $\gamma$-OL the $\gamma$-oligomer
is also stable with respect to the solution, but metastable with
respect to all ${\it i}\beta$-sheets. Finally, in region $\alpha$-OL,
though being stable with respect to the solution, the
$\alpha$-oligomer is metastable with respect to all other
aggregates. A similar hierarchy in the stability of protein aggregates
has been observed in an experimentally obtained phase diagram of Human
$\beta$B1-crystallin\cite{annunziata05}. The high stability of the
${\it i}\beta$-sheets arises from the fact that their structure is
largely due to the preponderant presence of hydrogen bonds between the
peptides\cite{knowles08}. Our finding of the existence of various
metastable peptide phases and of hierarchy in their metastability
provides a solid basis for describing the formation of amyloid fibrils
and for understanding why these fibrils can grow out of disordered
peptide aggregates\cite{serio00,auer08}.

%We thank Dr. Andrey Brukhno, Prof. Amos Maritan, Dr. Filip Meersman,
%Dr. Antonio Trovato and Dr. Wei-Feng Xue for enlightening
%discussions. 
One of the authors (D. K.) gratefully acknowledges the
financial support by the Leverhulme Trust (Grant F10100A) and the
hospitality that he enjoyed as Leverhulme Visiting Professor at the
University of Leeds.

%\bibliographystyle{revtex}
%\bibliography{biblio_title}

\end{document}